\documentclass[fleqn]{article}

\usepackage{spconf,amsmath,graphicx, enumitem}
\usepackage[acronym]{glossaries}
\makeglossaries
\newcommand{\comment}[1]{}

\title{Cell abundance aware deep learning for cell detection on highly imbalanced pathological data}
\name{%
	\begin{tabular}{@{}c@{}}
		Yeman Brhane Hagos$^{\star \dagger}$ \qquad 
		Catherine SY Lecat$^{\circ}$ \qquad 
		Dominic Patel $^{\diamond}$ \qquad 
		Lydia Lee $^{\circ}$ \qquad \\
		Thien-An Tran$^{\circ}$ \qquad 
		Manuel Rodriguez- Justo $^{\diamond}$ \qquad 
		Kwee Yong $^{\circ}$ \qquad 
		Yinyin Yuan$^{\star \dagger}$
\end{tabular}}
\address{$^{\star}$ Division of Molecular Pathology, The Institute of Cancer Research, London, UK. \\
	$^{\dagger}$Centre for Evolution and Cancer, The Institute of Cancer Research, London, UK. \\
	$^{\circ}$University College London Cancer Institute, Research Department of Haematology.\\
	$^{\diamond}$University College London Cancer Institute, Research Department of Pathology.\\
}

\newacronym{mihc}{mIHC}{Multiplex immunohistochemistry}
\begin{document}
	%
	\maketitle
	\begin{abstract}
		
		Automated analysis of tissue sections allows a better understanding of disease biology, and may reveal biomarkers that could guide prognosis or treatment selection. In digital pathology, less abundant cell types can be of biological significance, but their scarcity  can result in biased and sub-optimal cell detection model. To minimize the effect of cell imbalance on cell detection, we proposed a deep learning pipeline that considers the abundance of cell types during model training. Cell weight images were generated, which assign larger weights to less abundant cells and used the weights to regularize  Dice overlap loss function. The model was trained and evaluated on myeloma bone marrow trephine samples. Our model obtained cell detection F1-score of $0.78$, a $2\%$ increase compared to baseline models, and it outperformed baseline models at detecting rare cell types.  We found that scaling deep learning loss function by the abundance of cells improves cell detection performance. Our results demonstrate the importance of incorporating domain knowledge on deep learning methods for pathological data with class imbalance.
	\end{abstract}
	\section{Introduction}\label{sec:intro}
	
\begin{figure}[htb]
		\begin{minipage}[b]{1.0\linewidth}
			\centering
			\centerline{\includegraphics[width=6.0cm]{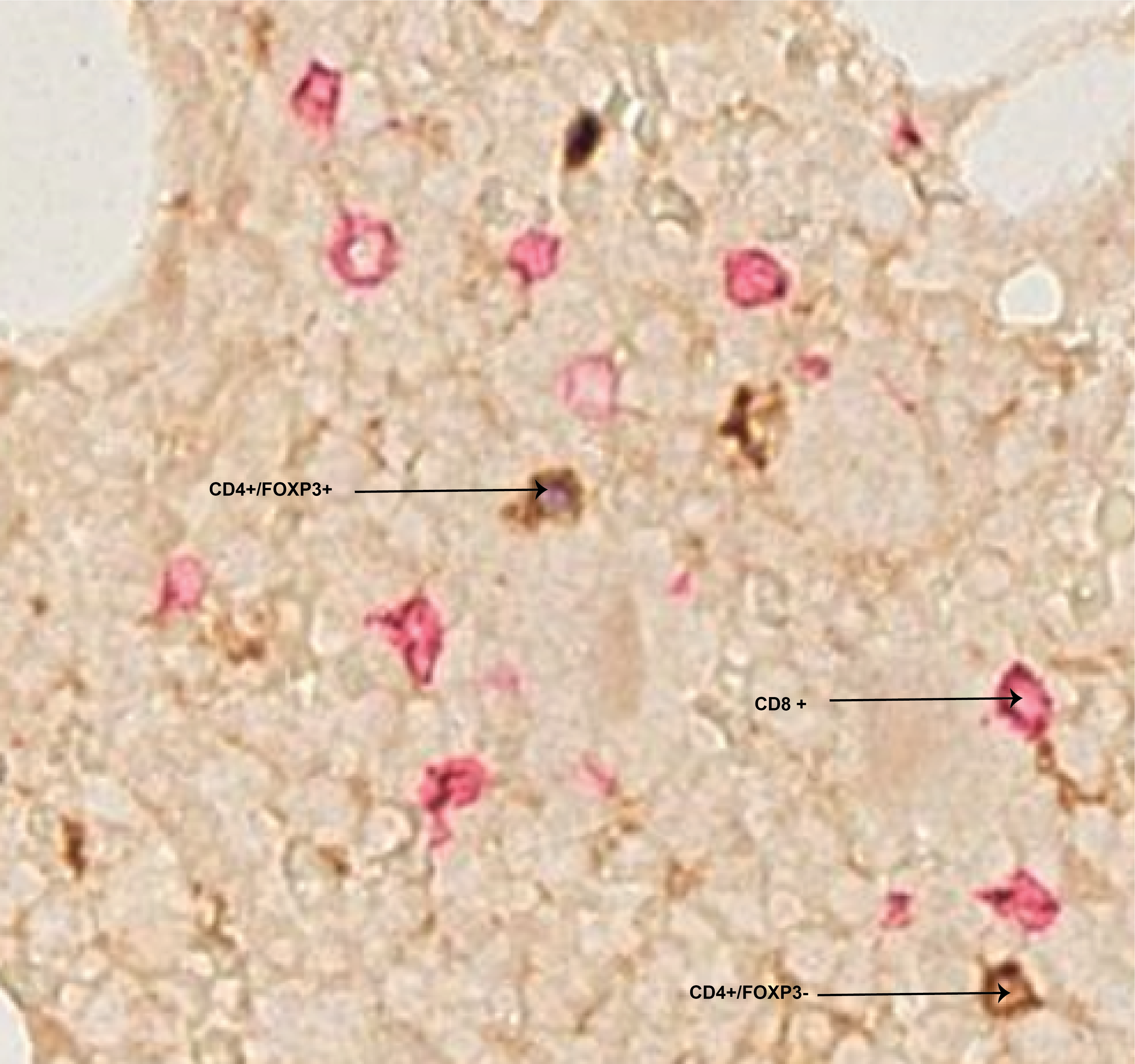}}
		\end{minipage}
		\caption{Samples mIHC image showing class imbalance. The number of CD8+ cells (red) are higher than CD4+/FOXP3- (brown) and CD4+/FOXP3+ (dark blue) cells.}
		\label{fig:fig1}
	\end{figure}

	In digital pathology, cell detection and classification are the first step to assessing tumour load, surrounding micro-environment and immune phenotypes \cite{yuan2016spatial}. \gls{mihc} is a staining method that allows simultaneous examination of multiple cell markers in a single image, where each cell is represented by a unique color or color combinations (Fig. \ref{fig:fig1}).  Intrinsically, some cell types are fewer compared to others. For example, in bone marrow trephine samples,  the number of CD4+/FOXP3- effector and CD4+/FOXP3+ regulatory T cells is lower than that of CD8+ T cells ( Fig. \ref{fig:fig1}). This imbalance causes instability and bias on the performance of discriminative models. 
	
	Recently,  different deep learning techniques have been proposed to address the issue of class imbalance in medical image data. Some methods focus on sampling, and/or augmentation \cite{sudre2017generalised}. The sampling method reduces variability of the data \cite{sudre2017generalised}. Both methods are suited for segmentation (for example, background vs. foreground segmentation) and classification applications  because a training sample for such applications has fewer number of instances/labels. However, in a patch based cell detection, there might be hundreds of cells in a small patch belonging to different classes. Thus, patch level sampling and augmentation approaches might increase the degree of imbalance in the context of single cell detection. Other methods focused on developing a robust training loss function \cite{sudre2017generalised,falk2019u}. Folk et al. \cite{falk2019u} proposed an approach that assigns cell weight from cell segmentation. However, collecting manual single cell segmentation is costly.
	
	In this work, we proposed a new class balancing approach in the context of single cell detection from single cell dot annotation. Our work has the following contributions:
	\begin{itemize}[topsep=0pt, partopsep=0pt]
		\item We implemented  cell detection and classification deep learning framework that uses class balancing technique of Dice overlap in dataset with class imbalance.
		\item We implemented an algorithm that generates cells weight image from expert dot annotation based on the relative abundance of cell types in the training data.
		\item We implemented and compared performance of different cell weighting strategies.
	\end{itemize}

	\section{Material}
	
	Our dataset contains $11$ newly diagnosed myeloma bone marrow \gls{mihc} whole slide images. It contains CD8+, CD4+/FOXP3- and CD4+/FOXP3+ cell types (Fig. \ref{fig:data}a). To train and evaluate the proposed method, a total of 8014 cells were annotated in different regions of the whole slide images by experts by putting a dot at the center of a cell (Fig. \ref{fig:data}b). Table \ref{tab:data} shows training, validation, and testing split.
	
	\begin{table} \label{tab:data}
		\begin{center}
			\caption{Distribution of dataset. number of slides ($N_{s}$)}
			\begin{tabular}{|l|c|c|c|c|}
				\hline
				-- & {\scriptsize { $\boldsymbol N_{s}$}} &  {\scriptsize \textbf{CD8+}} & {\scriptsize \textbf{\textbf{CD4+/FOXP3-}} } & {\scriptsize \textbf{CD4+/FOXP3+} }\\
				\hline
				Training &  5  & 2244 & 997 & 243\\
				\hline
				Validation &  3   & 1555 & 689 & 140\\
				\hline
				Test &  3  & 1306 & 702 & 138\\
				\hline
			\end{tabular}
		\end{center}
	\end{table}

	\section{Methodology}\label{sec:method}
	
	\subsection{Cell detection training data preparation}
	
\begin{figure}[htb]
		
		\begin{minipage}[b]{1.0\linewidth}
			\centering
			\centerline{\includegraphics[width=8.5cm]{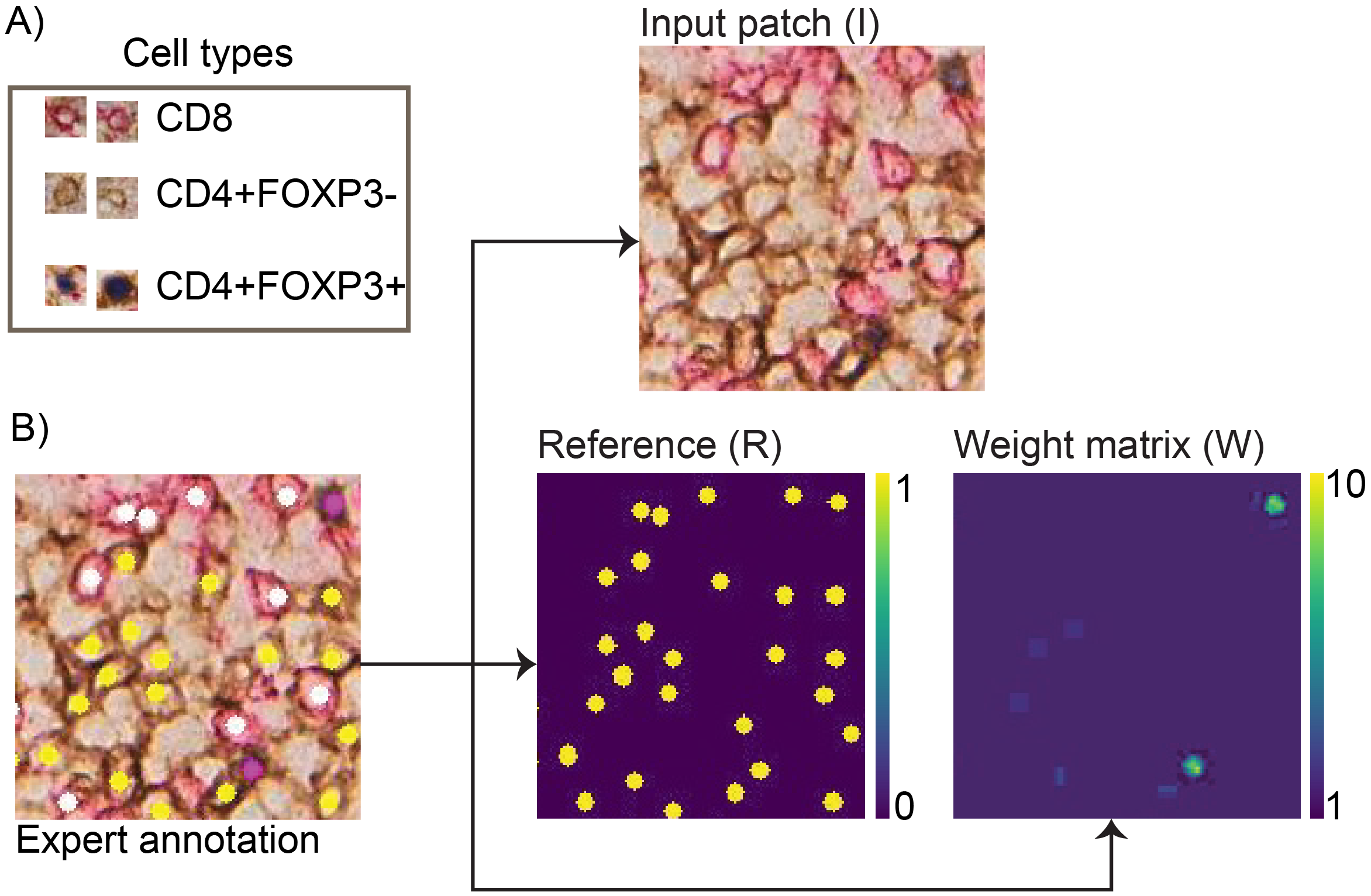}}
		\end{minipage}
		\caption{Training data preparation. a) Sample patches for all cell types. b). Sample annotated, reference (R) and weight (W) images for an input image (I). In W, less abundant cell type is assigned larger weight. CD4+/FOXP3+ cells have larger weight than CD4+/FOXP3- and CD8 cells.}
		\label{fig:data}
	\end{figure}

	For training, the annotated regions were divided into  $256 x 256 \newline x 3$ patches. Let $n$ be the number of training patches, the training data, $T_{d}$ is represented by a set 
	$T_{d} = \{\textbf{I}, \ \textbf{R}, \ \textbf{W})\} = \{(I^1, R^1, W^1),  (I^2,\   R^2, \ W^2), (I^3, \ R^3,\ W^3), ..., (I^n,  R^n, \newline W^n)\}$
	, where  $I^i \ \in \ R^{256 \ \text{x} \ 256\  \text{x} \  3}$, $W^i\ \in \ R^{256 \ \text{x} \ 256\  \text{x} \  1}$,   and $R^i\ \in \ R^{256 \ \text{x} \ 256\  \text{x} \  1}$ are the $i^{th}$ input, weight, and reference images, respectively. Sample I, R and W images are shown in Fig. \ref{fig:data}b.
	
	\textbf{Reference image: }it is an artificial image generated from the expert single cell dot annotation using Equation (\ref{eq:dot}).
	\begin{equation}\label{eq:dot}
	R(i, j) = \begin{cases} 
	1 & \text{if } d < r \\
	0       & \text{} otherwise
	\end{cases}
	\end{equation}
	where $R (i,j)$ is pixel value at $(i,j)$ and $d$ is an Euclidean distance from $(i, j)$ to the closest cell center. The value of $r$  was set to 4  pixels( 1.768 $\mu$ m). The value of $r$ was chosen empirically, making sure blobs in R don't touch each other (Fig. \ref{fig:data}b).
	
	\textbf{Weight image: } it assigns a weight to every cell in the input image (Fig. \ref{fig:data}b). The weights are inferred from the relative abundance of each cell type in the training data. Rare cells are given larger weight. Let $n$ be the number of cell types in the training dataset, let $C=\{c^1, c^2, c^3, ..., c^n \}$ be the $n$ cell types, and let $N^k$ be the number of $c^k$ cells in the training data. Then, $N=\{N^1, N^2, N^3, ..., N^n \}$ represent a set of abundance of cells for the $n$ cell types. Weight image is generated using information from $N$ and location of cells. We implemented three cell abundance weighting functions: RatioWeight (Equation \ref{eq:weight:linear}), ExpWeightType1  (Equation \ref{eq:weight:neg-exp}), and ExpWeightType2 (Equation \ref{eq:weight:neg-exp-sqr}).
	
	\begin{equation} \label{eq:weight:linear}
	W(i, j) = \begin{cases} 
	\frac{max(N)}{N^k} & \text{if } d^k < r \\
	1       & \text{} otherwise
	\end{cases}
	\end{equation}
	
	\begin{equation} \label{eq:weight:neg-exp}
	W(i, j) = \begin{cases} 
	\exp{-\frac{N^k}{max(N)}} & \text{if } d^k < r \\
	\exp{-1}       & \text{} otherwise
	\end{cases}
	\end{equation}
	
	\begin{equation} \label{eq:weight:neg-exp-sqr}
	W(i, j) = \begin{cases} 
	\exp{-(\frac{N^k}{max(N)})^2} & \text{if } d^k < r \\
	\exp{-1}       & \text{} otherwise
	\end{cases}
	\end{equation}
where $d^k$ is an Euclidean distance from $(i, j)$ to the $k^{th}$ cell type center. The value of $r$  was set to 4  pixels( 1.768 $\mu$m).

For our dataset, C = $\{$CD8+, CD4+/FOXP3-, CD4+/ FOXP3+ $\}$, N = \{$2244$, $997$, $243$\} and these values are used to generate the weights using Equation (\ref{eq:weight:linear} - \ref{eq:weight:neg-exp-sqr}).
	
	\subsection{Cell detection model}
	
		\begin{figure*}[ht]
		\begin{minipage}[b]{1.0\linewidth}
			\centering
			\centerline{\includegraphics[width=15.5cm]{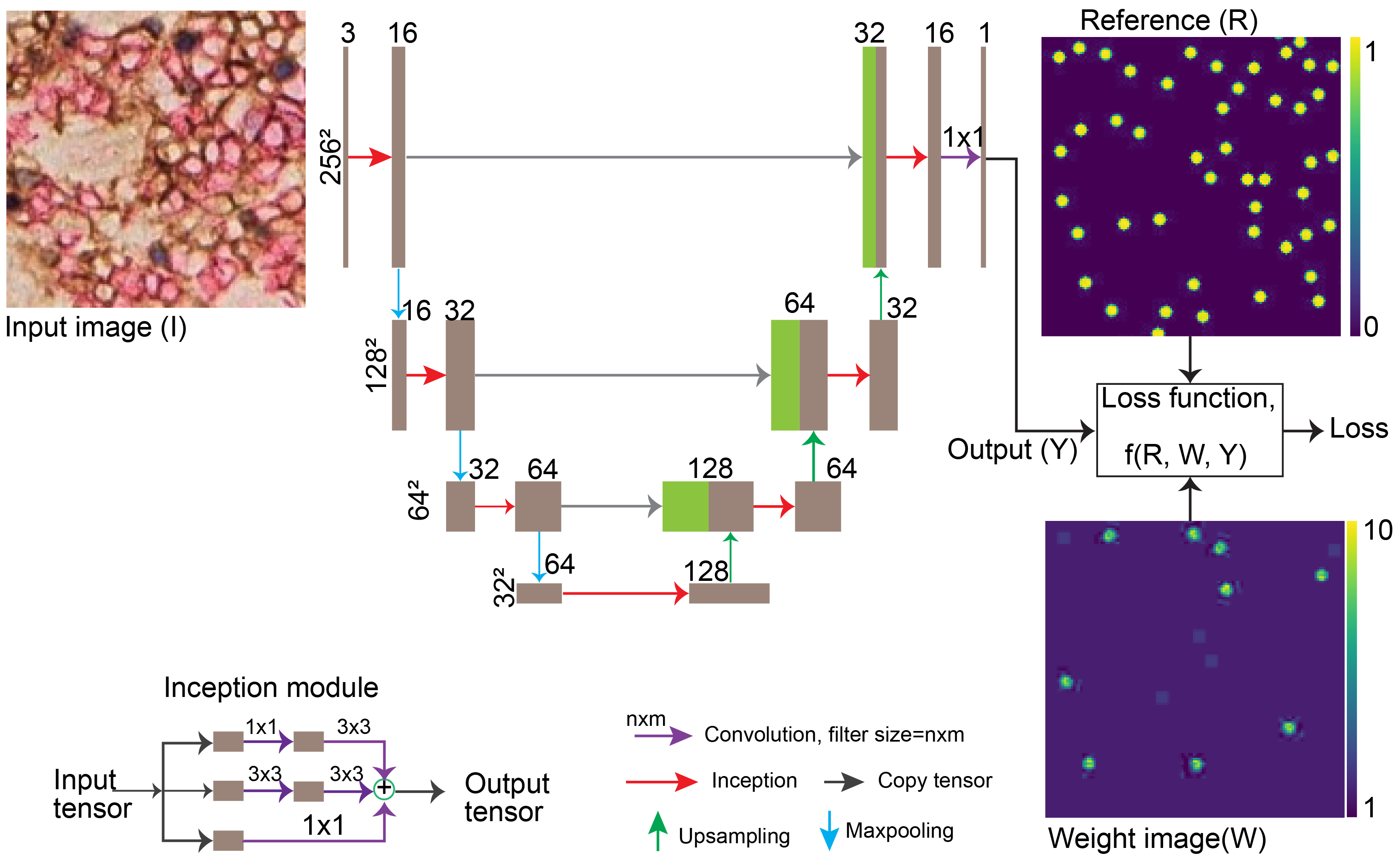}}
		\end{minipage}
		\caption{Schematic of cell detection model. The number on the top and side of the blocks indicate the size and spatial dimension of the features, respectively. The loss function is computed using model output, reference and weight images.}
		\label{fig:data-model}
	\end{figure*}

	Our proposed cell detection pipeline is shown in Fig. \ref{fig:data-model}. It is a U-net \cite{ronneberger2015u} convolutional neural network (CNN) inspired by inception V3. We applied inception V3 blocks which extracts multi-scale features at a given layer. The model has encoder and decoder part. The encoder learns low dimensional representation of the input image, and  the decoder reconstruct a target image. The $1x1$  convolutional layer at the end of the architecture transforms $256x256x16$ dimensional features to $256x256x1$, size of reference image (R). Parameters were initialized using uniform glorot \cite{glorot2010understanding}, and optimized using Adam \cite{kingma2014adam} with learning rate of $10^{-4}$. ReLU activation was applied to all layers, but Sigmoid to the last layer to transform the output to probability.
	
	\subsection{Cell detection loss function}
	
	To minimize the effect of class imbalance, we applied weighted dice overlap loss. The loss was computed as 
	\begin{equation}
	l = 	1 - 2 \frac{\sum_{j=0}^{j=w}\sum_{i=0}^{i=h}W(i, j)Y(i,j)R(i,j)\ + \  \epsilon}{\sum_{j=0}^{j=w}\sum_{i=0}^{i=h}W(i,j )(Y(i,j) + R(i,j))\ + \ \epsilon}
	\end{equation} where $W$, $R$ and $Y$ are the weight, reference and output images, respectively. $w$ and $h$ represent width and height of input image respectively. $\epsilon = 10^{-5}$ was added to ensure computational stability.
	
	\subsection{Cell classification model}
	
	To train a cell classification model, we extracted $28x28x3$ patches as shown in Fig. \ref{fig:data}a. We applied VGG \cite{simonyan2014very} style architecture, which contains three convolutional layers with  $\{ 16, 32, 64\}$ filters followed by two dense layers with  $\{ 200, 3\}$ neurons. ReLU activation was applied to all layer, but Softmax for the last layer to transform the tensors to probabilities. Parameters were initialized using uniform glorot \cite{glorot2010understanding}, and optimized using Adam \cite{kingma2014adam} with learning rate of $10^{-4}$. We applied categorical cross entropy loss with class weighting explained in Equation (\ref{eq:weight:linear} - \ref{eq:weight:neg-exp-sqr}).
	
	\section{Results and Discussion}\label{sec:result}
	
	To evaluate the performance of the proposed different weighting strategy based cell detection models and compare with other state of the art U-Net \cite{ronneberger2015u} and CONCORDe-Net \cite{hagos2019concorde}, we measured  precision, recall, and f1-score on separately held test images. CONCORDe-Net \cite{hagos2019concorde} is cell count regularized CNN designed for cell detection for mIHC images.
	
	F1-score of $0.78$ was obtained using ExpWeightType1 and RatioWeight models, a $2\%$ increase compared to U-net and $1\%$ increase compared to CONCORDe-Net \cite{hagos2019concorde}  as shown in Table \ref{tab:CDEval}. Moreover, recall of ExpWeightType2 model was higher than baseline models by at least $5\%$. For ExpWeightType1 model, a detection was considered as true positive if it is within $10$ pixels ($4.42\mu m$) Euclidean distance  to a ground truth annotation. For all models, the distance was optimized independently maximizing F1-score. This suggests class weighting improves cell detection performance.
	
\begin{figure}[htb]
		\begin{minipage}[b]{1.0\linewidth}
			\centering
			\centerline{\includegraphics[width=8.5cm]{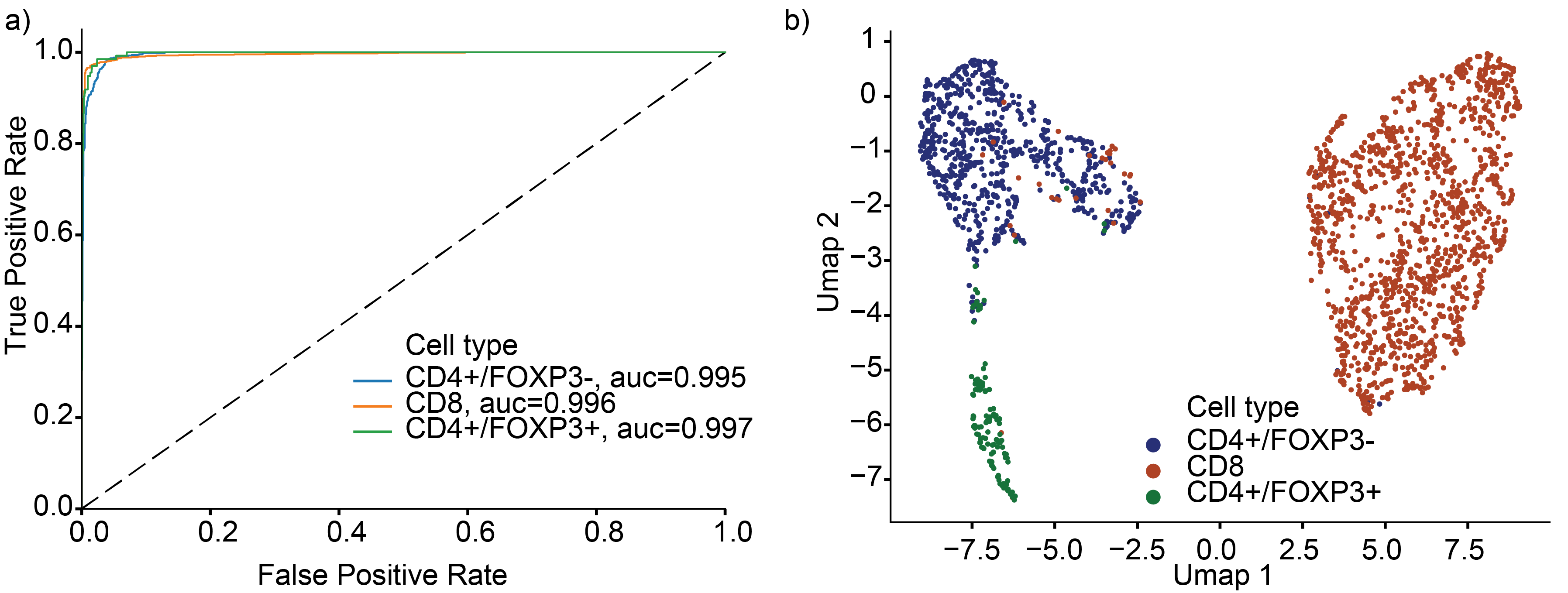}}
		\end{minipage}
		\caption{Cell classification model performance evaluation. a) Receiver operating characteristics curve and area under the curve (AUC). b) UMAP visualization of $200$ dimensional deep learned features.}
		\label{fig:cc_result}
	\end{figure}

	\begin{table}[h!]
		\begin{center}
			\caption{Cell detection performance of different models. U-net \cite{ronneberger2015u} model is a model in Fig. \ref{fig:data-model} trained without applying weights.}
			\label{tab:CDEval}
			\begin{tabular}{|l|c|c|c|}
				\hline
				\textbf{Method} &\textbf{Precision} &  \textbf{Recall} & \textbf{F1-score}\\
				\hline
				ExpWeightType1 &\textbf{ 0.82} & 0.75  & \textbf{0.78}\\
				\hline
				RatioWeight &  0.80  & 0.75  & 0.78 \\
				\hline
				ExpWeightType2 &  0.78 & \textbf{0.77}  & 0.77\\
				\hline
				CONCORDe-Net \cite{hagos2019concorde} &  0.81 & 0.72  & 0.76\\
				\hline
				U-net \cite{ronneberger2015u} &  0.80 & 0.70  & 0.76\\
				\hline
			\end{tabular}
		\end{center}
	\end{table}
To measure classification performance, we measured area under the curve (AUC) and accuracy on a separately held test images. For cell classification, there was no significant difference on AUC for the different weighting strategies. Fig. \ref{fig:cc_result}a shows the performance of cell classifier with ExpWeightType1 weighting. AUC  was greater than $0.995$ for all cell types. Overall accuracy of $0.97$ was achieved. 
	
To visualize separability of cell types using deep learnt features and to scrutinize miss-classified cell types, we applied uniform manifold approximation and projection (UMAP) dimensionality reduction (Figure \ref{fig:cc_result}b). The different cell types are mapped into different UMAP space in 2D. CD8 cells in the same space with CD4 cells are cells expressing both CD4 and CD8 proteins.
	
	\begin{figure}[htb]
		\begin{minipage}[b]{1.0\linewidth}
			\centering
			\centerline{\includegraphics[width=8.5cm]{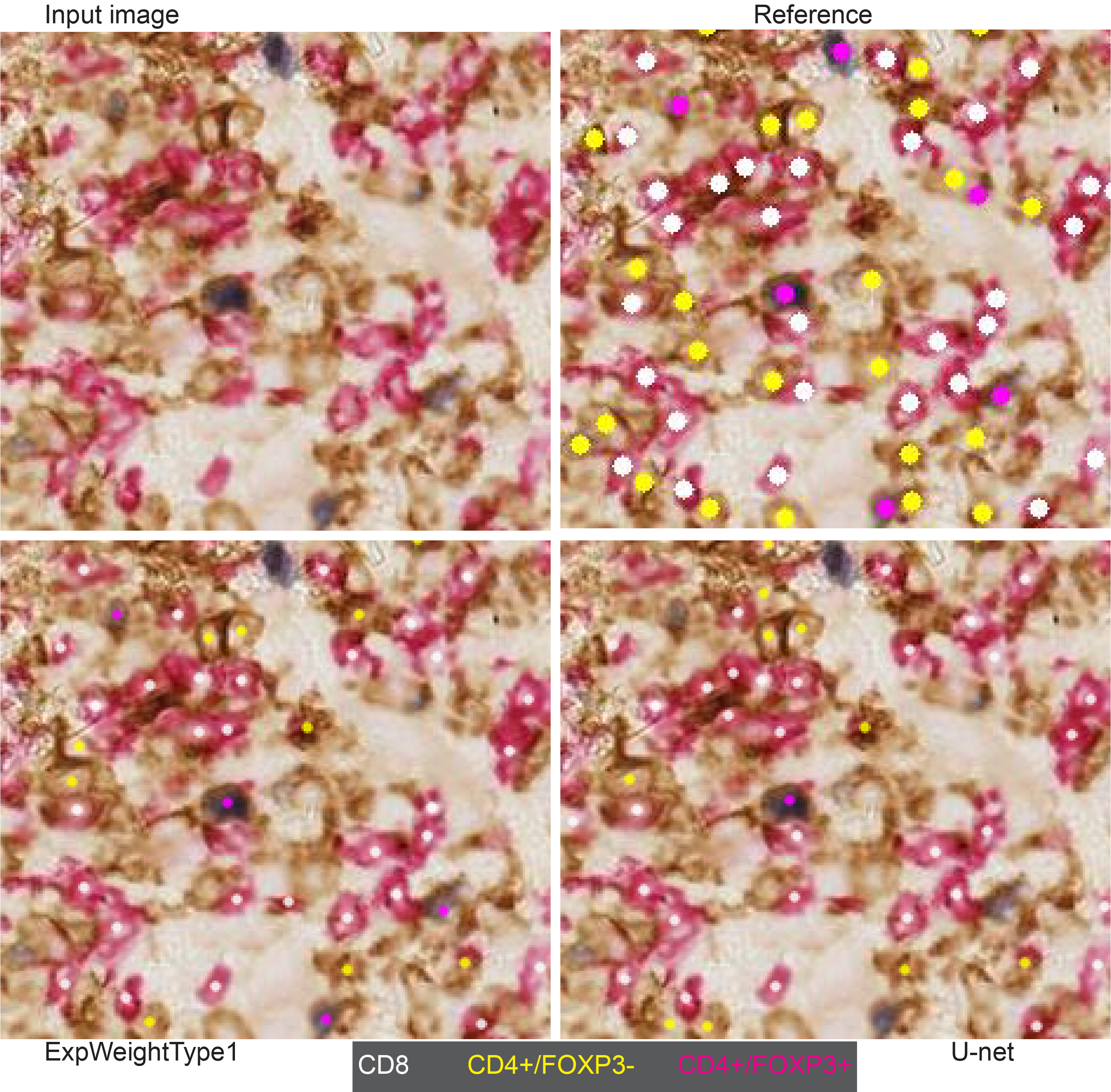}}
		\end{minipage}
		\caption{Samples results from different cell detection methods. }
		\label{fig:cc_cd_result}
	\end{figure}

	 In our dataset, compared to CD8+ cells there are less number of CD4+/FOXP3+ cells. The visualization in Fig. \ref{fig:cc_cd_result} indicates a model with ExpWeightType1 detected CD4+/FOXP3+ cells, which were under-represented in the training dataset, while the model trained without any weight (U-net) missed some of these cells. For CD8+ cells, the detection results remain similar with and without cell weighting. However, we observed  that the weighted model sometimes failed to detect large (in size) CD4+/FOXP3+ cells. This could be because of under-representation such type of cells in the training data. This indicates the proposed weighting method introduces as attention mechanism to the model to detect rare cell types and thus improve overall cell detection performance. 
	
	For reproducibility, model was trained using a Docker image within Singularity container on HPC cluster. Code is available at \textit{https://github.com/YemanBrhane/AwareNet}. The docker image is available on at \textit{yhagos/tf2gpu:concordenet} on Docker Hub.

	Our study has limitations. Our samples were collected from different hospitals with potential differences in processing and fixation, but they were stained and scanned using the same protocols and platform.  We trained and validated
	the model on small scale dataset.
	
	Overall, our results demonstrate the importance of incorporating domain knowledge for deep learning training on a dataset with class imbalance. In the future, we will apply the model on a larger cohort of bone marrow samples, to  understand the composition of the bone marrow immune microenvironment, and the changes imposed by malignant disease.
	
	\section{Conclusion}
	
	In this paper, to minimize the effect of cell imbalance in cell detection, we proposed a deep learning method that considers abundance of cells during training. Cell weight images were generated by assigning larger weights to less abundant cell types and applied the weights to regularize  Dice overlap loss function. Using negative exponential weighting, we obtained a $2\%$ increase in cell detection F1-score, and better rare cells detection compared to baseline models. 
	
	\section*{Compliance with Ethical Standards}
	
	The data used in this study was obtained with appropriate ethical approval granted by HRA and HCRW (REC Reference 07/Q0502/17).

	\section*{Acknowledgements}
	
	This project was funded by the European Union's Horizon 2020 research and innovation programme under the Marie Sklodowska-Curie grant agreement No766030 and CRUK Early Detection Program Award (C9203/A28770). KY receives funding from the National Institute for Health Research University College Hospital Biomedical Research Centre. LL is supported by the Medical Research Council, UK.
		
\bibliographystyle{IEEEbib}

\begin{thebibliography}{1}
	
	\bibitem{yuan2016spatial}
	Yinyin Yuan,
	\newblock ``Spatial heterogeneity in the tumor microenvironment,''
	\newblock {\em Cold Spring Harbor perspectives in medicine}, vol. 6, no. 8, pp.
	a026583, 2016.
	
	\bibitem{sudre2017generalised}
	Carole~H Sudre, Wenqi Li, Tom Vercauteren, Sebastien Ourselin, and M~Jorge
	Cardoso,
	\newblock ``Generalised dice overlap as a deep learning loss function for
	highly unbalanced segmentations,''
	\newblock in {\em Deep learning in medical image analysis and multimodal
		learning for clinical decision support}, pp. 240--248. Springer, 2017.
	
	\bibitem{falk2019u}
	Thorsten Falk, Dominic Mai, Robert Bensch, {\"O}zg{\"u}n {\c{C}}i{\c{c}}ek,
	Ahmed Abdulkadir, Yassine Marrakchi, Anton B{\"o}hm, Jan Deubner, Zoe
	J{\"a}ckel, Katharina Seiwald, et~al.,
	\newblock ``U-net: deep learning for cell counting, detection, and
	morphometry,''
	\newblock {\em Nature methods}, vol. 16, no. 1, pp. 67--70, 2019.
	
	\bibitem{ronneberger2015u}
	Olaf Ronneberger, Philipp Fischer, and Thomas Brox,
	\newblock ``U-net: Convolutional networks for biomedical image segmentation,''
	\newblock in {\em International Conference on Medical image computing and
		computer-assisted intervention}. Springer, 2015, pp. 234--241.
	
	\bibitem{glorot2010understanding}
	Xavier Glorot and Yoshua Bengio,
	\newblock ``Understanding the difficulty of training deep feedforward neural
	networks,''
	\newblock in {\em Proceedings of the thirteenth international conference on
		artificial intelligence and statistics}, 2010, pp. 249--256.
	
	\bibitem{kingma2014adam}
	Diederik~P Kingma and Jimmy Ba,
	\newblock ``Adam: A method for stochastic optimization,''
	\newblock {\em arXiv preprint arXiv:1412.6980}, 2014.
	
	\bibitem{simonyan2014very}
	Karen Simonyan and Andrew Zisserman,
	\newblock ``Very deep convolutional networks for large-scale image
	recognition,''
	\newblock {\em arXiv preprint arXiv:1409.1556}, 2014.
	
	\bibitem{hagos2019concorde}
	Yeman~Brhane Hagos, Priya~Lakshmi Narayanan, Ayse~U Akarca, Teresa Marafioti,
	and Yinyin Yuan,
	\newblock ``Concorde-net: Cell count regularized convolutional neural network
	for cell detection in multiplex immunohistochemistry images,''
	\newblock in {\em International Conference on Medical Image Computing and
		Computer-Assisted Intervention}. Springer, 2019, pp. 667--675.
	
\end{thebibliography}

\end{document}